\documentclass[aps,12pt,pra,showpacs,superscriptaddress,preprint]{revtex4}

\usepackage{amsmath}

\usepackage{subfig}

\usepackage{array}

\usepackage{amsfonts}

\usepackage{epsf}

\usepackage{epsfig}

\usepackage{amssymb}

\usepackage{bbm}

\usepackage{delarray}

\usepackage{caption}

\usepackage{amstext}

\usepackage{graphics}

\usepackage{epstopdf}

\usepackage{color}

\catcode`ð=\active

\defð{\u{g}}

\catcode`Ð=\active

\defÐ{\u{G}}

\catcode`Ý=\active

\defÝ{\. I}

\catcode`ö=\active

\defö{\"{o}}

\catcode`Ö=\active

\defÖ{\"O}

\catcode`ü=\active

\defü{\"{u}}

\catcode`Ü=\active

\defÜ{\"{U}}

\catcode`Þ=\active

\defÞ{\c{S}}

\catcode`þ=\active

\defþ{\c{s}}

\catcode`ý=\active

\defý{{\i}}

\catcode`ç=\active

\defç{\d{c}}

\catcode`Ç=\active

\defÇ{\d{C}}

\begin{document}

\title{\textbf{Non-Relativistic Phase Shifts via Laplace Transform Approach}}
\author{\small Altuð Arda}
\email[E-mail: ]{arda@hacettepe.edu.tr}\affiliation{Department of
Physics Education, Hacettepe University, 06800, Ankara,Turkey}
\author{\small Tapas Das}
\email[E-mail: ]{tapasd20@gmail.com}\affiliation{Kodalia Prasanna Banga High
 School (H.S), South 24 Parganas, Sonarpur 700146, India}

\begin{abstract}
The Laplace transform approach with convolution theorem is used to find the scattering phase shifts of a Mie-type potential. The normalized scattering wave functions are also studied. The bound state spectrum and the corresponding normalized wave functions are obtained with the help of the analytical properties of the scattering amplitude.
\end{abstract}


\maketitle

\newpage

\section{Introduction}

The theory within quantum mechanics plays an important role because it allows to predict experimental observations ruled by theory. Particularly, the scattering phase shifts are needed for a detailed analyse of interactions [1]. Among the computations of phase shifts for real potentials, the same subject has been also studied for potentials having complex form [1, 2]. One of these complex potentials has been proposed in Ref. [3] as an attractive Coulomb plus inverse-square potential with an imaginare constant
and complex phase shifts of the potential have been calculated in literature [1, 4]. This potential has a form of the Mie-type potential
\begin{eqnarray}
V(r)=-\frac{a_{1}}{r^{2}}-\frac{a_{2}}{r}+a_{3}\,,
\end{eqnarray}
which has been received a great attention in literature [5, references therein]. Arda and Sever have studied the bound state spectrum and the corresponding wave functions for this form of the potential by using the Laplace transform approach (LTA) [5]. The Mie-type potential gives the Kratzer-Fues potential [$V(r)=-D_e(\frac{2r_e}{r}-\frac{r_e^2}{r^2})$] if we chose the parameters as $a_{1}=-D_{e}r^{2}_{e}$, $a_{2}=2D_{e}r_{e}$, and $a_{3}=0$, while the modified Kratzer potential $[V(r)=-D_e\Big(\frac{r-r_e}{r}\Big)^2]$ is obtained by setting the parameters as $a_{1}=D_{e}r^{2}_{e}$, $a_{2}=-2D_{e}r_{e}$, and $a_{3}=D_{e}$ [6]. Here $D_e$ is the interaction energy between two atoms in a molecular system at equilibrium distance $r_0$. The potential in (1) serves also as a shifted, attractive Coulomb potential if we set $a_{1}=0$. In the present work, we extend the work for the case where the scattering phase shifts and normalized scattering wave functions will be studied by applying the same approach.

The present paper is arranged as follows. In Section 2, we collect the required points within LTA and give the convolution theorem briefly. In Section 3, we apply the above approach to find the exact scattering states and phase shifts of the Mie-type potential. We find also the normalized scattering wave functions. In Section 4, we study the analytical properties of the scattering amplitude, and find the energy eigenvalues with the help of it's poles. We write also the corresponding normalized wave functions of the bound states. We briefly discuss the similarity of the bounded results appearing in the non-relativistic Coulomb problem because of the existence of the Laplace-Runge-Lenz vector. We compare the radial form of the Schrödinger equation for the Mie-type potential with the Klein-Gordon equation for the charged particle in an external Coulomb field where the Sommerfeld parameter becomes slightly different. Finally, we give the conclusions in last Section.

\section{Laplace transform approach and Convolution Theorem}

Let us write a brief definition of the Laplace transform and collect a few of it's properties [7, 8]. The Laplace transform of a function $f(z)$ is given by
\begin{eqnarray}
F(s)=\mathcal{L}\left\{f(z)\right\}=\int_{0}^{\infty}e^{-{sz}}{f(z)}dz\,.
\end{eqnarray}

If there is some constant $\sigma\in \mathbb{R}$ such that ${\left|e^{-{\sigma}{z}}{f(z)}\right|\leq \bf M}$
for sufficiently large $z$, the integral in Eq. (2) will exist for  Re $s>\sigma$ . The Laplace transform may fail to exist because of a sufficiently strong singularity in the function $f(z)$ as $z \rightarrow 0$ . In particular
\begin{eqnarray}
\mathcal{L}\left\{\frac{z^{\alpha}}{\Gamma(\alpha+1)}\right\}=\frac{1}{s^{\alpha+1}}\,,{\alpha}>-1\,.
\end{eqnarray}
where $\Gamma(\alpha)$ is Gamma function. The followings are written for the Laplace transform of derivatives [7, 8]
\begin{eqnarray}
&&\mathcal{L}\left\{f^{(n)}(z)\right\}=s^n\mathcal{L}\left\{f(z)\right\}-\sum_{j=0}^{n-1}s^{n-1-j}{f^{(j)}(0)}\,,\nonumber\\
&&\mathcal{L}\left\{z^{n}f(z)\right\}=(-1)^{n}F^{(n)}(s)\,,
\end{eqnarray}
where the superscript $(n)$ denotes the $n$-th derivative with respect to $z$ for $f^{(n)}{(z)}$, and with respect to $s$ for $F^{(n)}{(s)}$.

According Eq. (2), the inverse Laplace transform $\mathcal{L}^{-1}$ is then defined as $\mathcal{L}^{-1}\left\{F(s)\right\}=f(z)$. The convolution theorem is related with inverse transform, and if we have two transformed function $g(s)=\mathcal{L}\left\{G(z)\right\}$ and
$h(s)=\mathcal{L}\left\{H(z)\right\}$, then the product of theme is the Laplace transform of the convolution $(G*H)(z)$
\begin{eqnarray}
(G*H)(z)=\int_{0}^z G(z-\tau)H(\tau)d\tau\,,
\end{eqnarray}
So the convolution theorem gives
\begin{eqnarray}
\mathcal{L}(G*H)(z)=g(s)h(s)\,,
\end{eqnarray}
and accordingly
\begin{eqnarray}
\mathcal{L}^{-1}\left\{g(s)h(s)\right\}=\int_{0}^z G(z-\tau)H(\tau)d\tau\,.
\end{eqnarray}
If we substitute $w=z-\tau$, then we find the important consequence $G*H=H*G$. Lastly, the following equalities for inverse Laplace transform which will be used in below are required for the computation [7, 8]
\begin{eqnarray}
\mathcal{L}^{-1}\left\{(s+\beta)^{-\,a}\right\}=G(z)=\frac{z^{a-1}{e^{-\beta z}}}{\Gamma(a)}\,,\nonumber\\
\mathcal{L}^{-1}\left\{(s-\beta)^{-\,b}\right\}=H(z)=\frac{z^{b-1}{e^{\beta z}}}{\Gamma(b)}\,.
\end{eqnarray}
where $a, b$, and $\beta$ are some constant parameters.

\section{Exact Scattering States and Phase Shifts}

In spherical coordinates $(r, \theta, \phi)$, the Schrödinger equation with energy $E$ is given by ($e=\hbar=m=1$)
\begin{eqnarray}
\left[-\frac{1}{2}\,\vec{\nabla}^2+V(r)\right]\Psi(r, \theta, \phi)=E\Psi(r, \theta, \phi)\,,
\end{eqnarray}
where the wave function $\Psi(r, \theta, \phi)$ for a spherical symmetric potential is written as $\Psi(r, \theta, \phi)=\frac{1}{r}R(r)Y(\theta, \phi)$ with the spherical harmonics $Y(\theta, \phi)$. By inserting the potential in (1) into Eq. (9), we obtain
\begin{eqnarray}
\frac{d^2 R(r)}{dr^2}+\left[\frac{2a_{2}}{r}-\frac{\ell(\ell+1)-2a_{1}}{r^2}+\varepsilon^2\right]R(r)=0\,.
\end{eqnarray}
where $\varepsilon^2=2(E-a_{3})$, and $\ell$ is the orbital angular momentum quantum number. Here it is interesting to note down that the above radial equation is equivalent to the stationary Klein-Gordon equation for charged particle in external Coulomb potential. Let us verify the point very briefly in usual convention of unit. The K-G particle of mass $m$ within the interaction of electromagnetic field is written as
\begin{eqnarray*}
\Big(i\hbar\frac{\partial}{\partial t}-eV\Big)^2\Psi(\vec{x},t)-c^2\Big(-i\hbar\vec{\nabla}-\frac{e\vec{A}}{c}\Big)^2\Psi(\vec{x},t)-m^2c^4\Psi(\vec{x},t)=0\,,
\end{eqnarray*}
where $\vec{A}$ and $V$ represent the vector and scalar potential of the field respectively. $c, e$ are the velocity of light and charge of electron. For Coulomb potential $\vec{A}=0$ and $eV=-\frac{Ze^2}{r}$ i.e potentials are time independent. So letting the solution as $\Psi(\vec{x},t)=u(\vec{x})e^{-iEt}$ we can have
\begin{eqnarray*}
\Big(E+\frac{Ze^2}{r}\Big)^2u=(-\hbar^2 c^2\vec{\nabla}^2+m^2c^4)u\,.
\end{eqnarray*}
Now since the potential is spherically symmetric, we can take $u(\vec{x})=\frac{1}{r}R(r)Y_{\ell m}(\theta,\phi)$ and this immediately  provides the radial equation (via the angular part's eigenvalue equation) as
\begin{eqnarray*}
\bigg[\frac{d^2R}{dr^2}+\frac{E^2-m^2c^4}{\hbar^2 c^2}+\frac{2EZe^2}{\hbar^2 c^2 r}-\frac{\ell(\ell+1)-Z^2e^4/{\hbar^2 c^2}}{r^2}\Bigg]R=0\,.
\end{eqnarray*}
Taking natural unit (for mathematical convenience) we can say that equation is clearly similar to the Eq.(10).  Now proceeding further to the our problem we know that the boundary conditions on the solutions of Eq. (10) is $R(r \rightarrow 0)=0$, and $R(r \rightarrow \infty)$ is a finite value. Following the above boundary condition, we use the predetermined ansatz for the wave functions
\begin{eqnarray}
R(r)=Nr^{A+\frac{1}{2}}e^{-i\varepsilon r}f(r)\,.
\end{eqnarray}
where $i=\sqrt{-1\,}$. Substituting it into Eq. (10), and using a new variable as $z=-2i\varepsilon r$, we have
\begin{eqnarray}
z\frac{d^2f(z)}{dz^2}+(2A+1+z)\frac{df(z)}{dz}+\left(A+\frac{1}{2}+\frac{ia_{2}}{\varepsilon}\right)f(z)=0\,,
\end{eqnarray}
with $A^2=\ell(\ell+1)-2a_{1}+1/4$.

Using the LTA defined in Eq. (2) turns Eq. (12) to following first-order differential equation
\begin{eqnarray}
s(s+1)\frac{dF(s)}{ds}-[s(2A-1)+A-\frac{1}{2}+\frac{ia_{2}}{\varepsilon}]F(s)=0\,,
\end{eqnarray}
for which the solution can be written as
\begin{eqnarray}
F(s)=(s+1)^{-(A+\frac{1}{2}+\frac{ia_{2}}{\varepsilon})}s^{-(A+\frac{1}{2}-\frac{ia_{2}}{\varepsilon})}=Ng(s)h(s)\,.
\end{eqnarray}
where $N$ is a normalization constant obtained below. In order to find the solutions of Eq. (12), we use Eqs. (7)-(8), and write
\begin{eqnarray}
f(z)&=&\mathcal{L}^{-1}\left\{F(s)\right\}=N(G*H)(z)=N\int_{0}^z G(z-\tau)H(\tau)d\tau\nonumber\\
&=&\frac{N}{\Gamma(-A+\frac{1}{2}-\frac{ia_{2}}{\varepsilon})\Gamma(-A+\frac{1}{2}+\frac{ia_{2}}{\varepsilon})}
\int_{0}^z (z-\tau)^{(-A-\frac{1}{2}-\frac{ia_{2}}{\varepsilon})}\,\tau^{(-A-\frac{1}{2}+\frac{ia_{2}}{\varepsilon})} e^{-\tau}d\tau\,.
\end{eqnarray}

With the help of the following formula [9]
\begin{eqnarray}
\int_{0}^y (y-\tau)^{a'-1}\tau^{b'-1} e^{\mu\tau}d\tau=\frac{\Gamma(a')\Gamma(b')}{\Gamma(a'+b')}y^{a'+b'-1}\,_{1}F_{1}(b', a'+b', \mu y)\,,
\end{eqnarray}
and the property satisfying by the confluent hypergeometric functions $\,_{1}F_1(p;q;z)$ as $\,_{1}F_1(p;q;-z)=e^{-z}\,_{1}F_{1}(q-p;q;z)$ [10],
we write the scattering state wave functions for Mie-type potential as
\begin{eqnarray}
f(z)=\frac{N}{\Gamma(1-2A)}z^{-2A}e^{-z}\,_{1}F_{1}(-A+\frac{1}{2}+\frac{ia_{2}}{\varepsilon};-2A+1;z)\,.
\end{eqnarray}

We are now in a position to study the asymptotic form of the above expression, and find the normalization constant and phase shifts. The asymptotic form of the confluent hypergeometric function for $|z| \rightarrow \infty$ [9] is
\begin{eqnarray}
\,_{1}F_{1}(\omega_1;\omega_2;z) \rightarrow \frac{\Gamma(\omega_2)}{\Gamma(\omega_1)}\,e^{z}z^{\omega_1-\omega_2}+\frac{\Gamma(\omega_2)}{\Gamma(\omega_2-\omega_1)}e^{\pm i\pi\omega_1/2}z^{-\omega_1}\,,
\end{eqnarray}
where the upper sign in second term corresponds to $-\pi/2<arg z<3\pi/2$, while the other sign corresponds to $-3\pi/2<arg z<\pi/2$. The last equality is written for $z=-2i\varepsilon r=|z|e^{-i\pi/2}$ as
\begin{eqnarray}
\,_{1}F_{1}(\omega_1;\omega_2;z) \rightarrow \frac{\Gamma(\omega_2)}{\Gamma(\omega_1)}\,e^{z}|z|^{\omega_1-\omega_2}e^{-i\pi(\omega_1-\omega_2)/2}+
\frac{\Gamma(\omega_2)}{\Gamma(\omega_2-\omega_1)}e^{-i\pi\omega_1/2}z^{-\omega_1}\,.
\end{eqnarray}
With the help of this equation, the confluent hypergeometric function in Eq. (17) is given by
\begin{eqnarray}
&&\,_{1}F_{1}(-A+1/2-ia_{2}/\varepsilon;1-2A;-2i\varepsilon r)\nonumber\\ && \xrightarrow[r \to \infty]{} \frac{\Gamma(1-2A)}{\Gamma(-A+1/2-ia_{2}/\varepsilon)}e^{-2i\varepsilon r}(2\varepsilon r)^{A-1/2-ia_{2}/\varepsilon}
e^{-i\pi(A-1/2-ia_{2}/\varepsilon)/2}\nonumber\\&&+ \frac{\Gamma(1-2A)}{\Gamma(-A+1/2+ia_{2}/\varepsilon)}(2\varepsilon r)^{-(-A+1/2-ia_{2}/\varepsilon)}e^{-i\pi(-A+1/2-ia_{2}/\varepsilon)/2}\,.
\end{eqnarray}

If writing $\Gamma(-A+1/2-ia_{2}/\varepsilon)=|\Gamma(-A+1/2-ia_{2}/\varepsilon)|e^{i\delta'}$ then we have $\Gamma(-A+1/2+ia_{2}/\varepsilon)=|\Gamma(-A+1/2-ia_{2}/\varepsilon)|e^{-i\delta'}$ with a real number $\delta'$ which will correspond to phase shift. By using these equalities, we obtain from Eq. (20)
\begin{eqnarray}
&&\,_{1}F_{1}(-A+1/2-ia_{2}/\varepsilon;1-2A;-2i\varepsilon r)  \xrightarrow[r \to \infty]{} \frac{\Gamma(1-2A)}{|\Gamma(-A+1/2-ia_{2}/\varepsilon)|}\,e^{-i\varepsilon r}(2\varepsilon r)^{A-1/2}e^{-\pi a_{2}/2\varepsilon}\nonumber\\&\times&\left[ie^{-i[\delta'+\varepsilon r+(a_{2}/\varepsilon)ln(2\varepsilon r)+\frac{\pi}{2}(A+1/2)]}-ie^{i[\delta'+\varepsilon r+(a_{2}/\varepsilon)ln(2\varepsilon r)+\frac{\pi}{2}(A+1/2)]}\right]\,.
\end{eqnarray}
Finally, by substituting Eq. (21) into Eq. (17), we get from Eq. (11)
\begin{eqnarray}
R(r) \xrightarrow[r \to \infty]{} N\frac{(2\varepsilon)^{-1/2}e^{-\pi a_{2}/2\varepsilon}}{|\Gamma(-A+1/2-ia_{2}/\varepsilon)|} 2\sin (\delta'+\varepsilon r+\frac{a_{2}}{\varepsilon}\,ln(2\varepsilon r)+\frac{\pi}{2}(A+1/2))\,.
\end{eqnarray}
Here we will compare the basic scattering state wave function with what we have obtained till now. The basic Coulomb scattering state is characterized by a useful dimensionless parameter which is called as Sommerfeld parameter. The parameter is defined as $\eta=\frac{Z_1Z_2e^2}{\hbar v}=\alpha Z_1Z_2\sqrt{\frac{\mu c^2}{2E}}$, where by definition it is the situation of the collision of two particles with respective masses $m_1$ and $m_2$ and charges $Z_1e$ and $Z_2e$ at a positive energy $E$. $v=\hbar k/\mu$ is the relative velocity of the particles in the centre of mass frame. The wave number is defined as $k=\frac{\sqrt{2\mu E}}{\hbar}$  where $\mu=\frac{m_1m_2}{m_1+m_2}$ is the reduced mass. $\alpha=\frac{e^2}{4\pi\epsilon_0 \hbar c}$ is the fine structure constant. Now selecting the unit $\mu=1,\hbar=1, c=1, e=1$ one can easily find the Coulomb potential $V(r)=\frac{\eta k}{r}$. Clearly neutral case of scattering is recovered by $\eta=0$. \\ For a central potential, a wave function can be factorized in spherical coordinates $\vec{r}=r(r,\Omega)$ as $\psi(\vec{r})=\frac{1}{r}R(r)Y_{\ell}^m(\Omega)$. The spherical harmonics $Y_{\ell}^m(\Omega)$ depend on the orbital quantum number $\ell$ and magnetic quantum number $m$ as well as on the angles $\Omega=(\theta,\phi)$. The radial Schr\"{o}dinger equation for the Coulomb potential in partial wave $\ell$ reads
\begin{eqnarray*}
\Big(\frac{d^2}{dr^2}-\frac{\ell(\ell+1)}{r^2}-\frac{2k\eta}{r}+k^2\Big)R(r)=0\,.
\end{eqnarray*}
The wave function for the scattering states is written
\begin{eqnarray}
R(r) \xrightarrow[r \to \infty]{} 2\sin (kr+\delta-\frac{\pi}{2}\,\ell+\eta ln(2kr))\,,
\end{eqnarray}
then the radial wave functions of scattering states for Coulomb potential in non-relativistic case are normalized on the "$k/2\pi$" scale [10, 11]. Here $\delta$ represents phase shifts. Comparing Eq. (22) with Eq. (23) gives us the normalization constant of scattering states
\begin{eqnarray}
N=\sqrt{2\varepsilon\,}e^{\pi a_{2}/2\varepsilon}|\Gamma(-A+1/2-ia_{2}/\varepsilon)|\,,
\end{eqnarray}
and the phase shifts
\begin{eqnarray}
\delta=\delta'+\frac{\pi}{2}\left(A+\ell+\frac{1}{2}\right)=arg\Gamma(-A+1/2-ia_{2}/\varepsilon)+\frac{\pi}{2}\left(A+\ell+\frac{1}{2}\right)\,.
\end{eqnarray}
Inserting Eq. (24) into Eq. (11), we obtain the normalized wave functions of scattering states as
\begin{eqnarray}
R(r)=\frac{|\Gamma(-A+1/2-ia_{2}/\varepsilon)|e^{\pi a_{2}/2\varepsilon}}{(2\varepsilon)^{2A-(1/2)}\Gamma(1-2A)}\,r^{-A+\frac{1}{2}}e^{i\varepsilon r}\,_{1}F_{1}(-A+1/2-ia_{2}/\varepsilon;1-2A;-2i\varepsilon r)\,.
\end{eqnarray}
We study the analytical properties of the scattering amplitude to find the energy levels of the potential in next section.

\section{Scattering Amplitude and Analytical Properties}

According to the partial-wave method, the scattering amplitude is given by [11]
\begin{eqnarray}
F(\theta)=\sum_{\ell=0}^{\infty}(2\ell+1)\left[\frac{e^{2i\delta}}{2ik}\right]P_{\ell}(\cos\theta)\,,
\end{eqnarray}
where $\ell$ is the angular quantum number. In order to discuss the analytical properties of the scattering amplitude, we have to study the analytical properties of $\Gamma(-A+1/2-ia_{2}/\varepsilon)$ which can be done by using the definition of Gamma function
\begin{eqnarray}
\Gamma(z)=\frac{\Gamma(z+1)}{z}=\frac{\Gamma(z+2)}{z(z+1)}=\frac{\Gamma(z+3)}{z(z+1)(z+2)}=\ldots\,,
\end{eqnarray}
which means that the Gamma function has poles at $z=0, -1, -2, \ldots$ in the complex plane. Namely, the first-order poles of $\Gamma(-A+1/2-ia_{2}/\varepsilon)$ are situated at $-A+1/2-ia_{2}/\varepsilon=0, -1, -2, \ldots=-n$, $n=0, 1, 2, \ldots$. At the poles of phase shifts, the energy levels are given by
\begin{eqnarray}
E=a_{3}-\frac{a^{2}_{2}/2}{\left[n+\frac{1}{2}+\sqrt{\ell(\ell+1)+\frac{1}{4}-2a_{1}\,}\right]^2}\,,
\end{eqnarray}
which is exactly the energy equation of the bound states [5]. The Mie-type potential includes an "extra" term to the Coulomb potential, which modifies the centrifugal barrier. This term breaks the degeneracy in the Coulomb problem for the Schrödinger equation between levels of the same principal but different orbital angular momentum quantum number. The similar situation appears in the non-relativistic Coulomb problem because of the extra "accidental" degeneracy in the energy spectrum, which is a result of the existence of an extra operator called as Laplace-Runge-Lentz vector that commutes with the Hamiltonian.

By using the above equality $-A+1/2-ia_{2}/\varepsilon=-n$, the radial wave function given in Eq. (26) can be written as
\begin{eqnarray}
R(r)=N'r^{-A+\frac{1}{2}}\,e^{-a_{2}r/(n+1/2-A)}\,_{1}F_{1}\left(-n;1-2A;\frac{2a_{2}}{n+\frac{1}{2}-A}\,r\right)
\end{eqnarray}
which should be satisfied the normalization condition $\int_{0}^{\infty}|R(r)|^2r^2dr=1$. In order to find the normalization constant, we use the relation between the confluent hypergeometric functions and the generalized Laguerre polynomials [9]
\begin{eqnarray}
\,_{1}F_{1}(-n;\kappa+1;y)=\frac{n!\Gamma(\kappa+1)}{\Gamma(n+\kappa+1)}\,L_{n}^{\kappa}(y)\,,
\end{eqnarray}
and the following recurrence relation between the generalized Laguerre polynomials [12]
\begin{eqnarray}
yL_{n}^{\kappa}(y)=(\kappa+2n+1)L_{n}^{\kappa}(y)-(n+\kappa)L_{n-1}^{\kappa}(y)-(n+1)L_{n+1}^{\kappa}(y)\,,
\end{eqnarray}

We find the normalization constant for the bound state solutions as
\begin{eqnarray}
N'=\frac{(2a_{2})^{1-A}}{\Gamma(1-2A)}\,\sqrt{\frac{\Gamma(n+1-2A)}{2n!}\left(n+\frac{1}{2}-A\right)^{2A-3}\,}\,,
\end{eqnarray}
where used the orthogonality relation of the generalized Laguerre polynomials [9]
\begin{eqnarray}
\int_{0}^{\infty}y^{\eta}e^{-y}\left[L_{n}^{\eta}(y)\right]^2dy=\frac{\Gamma(\eta+n+1)}{n!}\,.
\end{eqnarray}

Finally we obtain the normalized radial wave functions for the Mie-type potential as
\begin{eqnarray}
R(r)=\sqrt{\frac{n!}{\Gamma(n+1-2A)}\,}\frac{1}{n+\frac{1}{2}-A}r^{-A+\frac{1}{2}}e^{-\frac{2a_{2}}{n+1/2-A}\,r}\,L_{n}^{-2A}\left(\frac{2a_{2}}{n+\frac{1}{2}-A}\,r\right)\,.
\end{eqnarray}
which is consistent with the earlier results obtained in literature [5].

\section{Conclusions}

We have analyzed the exact scattering states of the Schrödinger equation for a Mie-type potential which means that the normalized radial wave functions of scattering states, and the phase shifts have been obtained. For this aim, we have used the Laplace transform approach with the convolution theorem, and observed that this approach could be an economical path to analyze the problem. By discussing the analytical properties of scattering amplitude, we have found the energy levels of bound states, and the corresponding normalized radial wave functions. It is observed that the energy equation at the poles of the scattering amplitude corresponds to energy equation of the bound states, and radial wave functions of scattering states correspond to the radial wave functions of the bound states. It is found that the results obtained for the bound states are consistent with the results given in literature. We have discussed the similarity between our results obtained for the bounded part and the ones appearing in the non-relativistic Coulomb problem because of the Laplace-Runge-Lenz vector. We have also compared the radial form of the Schrödinger equation for the Mie-type potential with the Klein-Gordon equation for a charged particle in an external Coulomb potential including the Sommerfeld parameter.

\section{Acknowledgments}
The authors would like to thank the editor and the
anonymous reviewer for valuable comments. This research was supported by Hacettepe University Scientific Research Coordination Unit, Project Code: FBB-$2016$-$9394$.

\end{document}